########################Paper starts####################################
\documentstyle[12pt]{article}
\def\be{\begin{equation}}
\def\ee{\end{equation}}
\def\ba{\begin{array}}
\def\ea{\end{array}}
\def\beqn{\begin{eqnarray}}
\def\eeqn{\end{eqnarray}}

\title{Fritzsch-Xing mass matrices, $V_{td}$ and CP Violating phase
$\delta$}
\author{P.S.Gill$^*$  and Manmohan Gupta \\
{\it  Department of Physics, Panjab University, Chandigarh-160014, India. } }
\date{}
\begin{document}
\maketitle
\begin{abstract}
Natural $4$ zeros texture mass matrices recently proposed by Fritzsch and
Xing have been investigated by including `non-leading'corrections in the
context of latest data regarding $m_t^{pole}$ and $V_{CKM}$ matrix
elements. Apart from accommodating $m_t^{pole}$ in the range $175\pm15$
GeV, $|V_{cb}|$ and $|V_{ub}/V_{cb}|=0.08\pm0.02$, the analysis with
maximal CP-violation predicts $|V_{td}| = .005-.013$.  Further, the 
CP-violating phase angle $\delta$ can be restricted to the ranges ({\rm
i}) $22^o$ -$45^o$ and ({\rm ii}) $95^o$ - $130^o$, concretizing the
ambiguity regarding phase of CKM matrix.  Furthermore, we find that
non-leading calculations are important when `Cabibbo triangle' is to be
linked to unitarity triangle.
\end{abstract}
\vskip 5truecm
$^{\ast}$Permanent Address: S.G.G.S.College, Chandigarh~-~160026.
\newpage
Recently, Peccei and Wang \cite{r1} in a very interesting paper have found
the possible pattern of natural mass matrices at the GUT scale which are
in agreement with low energy data related to CKM mixing matrix. A concrete
realization of such mass matrices at low energy scale as shown by Wang
\cite{r2} is presented by the Fritzsch and Xing (FX) ansatz \cite{r3}
consisting of $4$ zeros texture mass matrices \cite{r4}. Exploiting the
idea of maximality of CP violation and relating the usual unitarity
triangle with `Cabibbo triangle', they have found some very interesting
results at the leading order. In particular, by adding non-zero
$22$-elements in the `U' as well as `D' sector to the usual Fritzsch mass
matrices \cite{r5}, FX have found at the leading order
\be V_{us} = \sqrt{\frac{m_d} {m_s }} - \sqrt{\frac{m_u}{m_c}} e^{i \Delta
\sigma},\label{e1} \ee 
\be V_{cd} =\sqrt{\frac{m_u} {m_c}} -
\sqrt{\frac{m_d}{m_s}}e^{i\Delta\sigma},\label{e2}\ee 
\beqn
\frac{V_{ub}}{V_{cb}} \approx \sqrt\frac{m_u}{m_c } & {\em and } & 
\frac{V_{td}}{V_{ts}} \approx \sqrt{\frac{m_d}{m_s}}. \label{e3} 
\eeqn 
Further, in the complex plane by linking the Cabibbo triangle with the
usual unitarity triangle, they have been able to show $\Delta\sigma
\approx 90^o$, implying maximal CP-violation in the context of present
mass matrices. The value of $\Delta\sigma$ is sensitive to variations of
mass ratios , $m_u /m_c$ and $m_d/m_s$, however, the above conclusion
about $\Delta\sigma$ is not inconsistent with the range suggested by such
variations.\\ 
     The purpose of the present brief report, on the one hand , is
to find non-leading order corrections to relations (\ref{e1})-(\ref{e3}),
on the other hand we want to examine the detailed implications of extra
terms, introduced by FX in their formalism compared to earlier Fritzsch
mass matrices \cite{r5}, on the $V_{CKM}$ phenomenology. Further, to
extend the success of FX mass matrices, it becomes interesting to examine
how FX mass matrices accommodate $m^{pole}_t$ \cite{r6}, latest data
regarding $V_{CKM}$ elements $V_{ub}$ and $V_{cb}$ as well as the recently
found range of $V_{td}$ due to improved QCD calculations of $B^0$ -
$\bar{B^0}$ mixing phenomenon \cite{r7}.  Furthermore, it would be
worthwhile to study the implications of non-leading corrections to 
angles of the unitarity triangle which may shed some light on  the
ambiguity relating to the phase of the CKM mixing matrix \cite{r8}.\\
To this end, we have first exactly diagonalized FX matrices and calculated
the corresponding $V_{CKM}$ . The implications of the extra terms both in
`U' and `D' sectors is quite manifest in the expressions for $V_{CKM}$
matrix elements derived here.\\ 
To begin with, we consider Fritzsch-Xing mass matrices \cite{r3}, for
example, 
\be M_i = \left( \ba{lll} 0 & D_i & 0\\
D_i^{\ast} & C_i & B_i\\ 0 & B_i & A_i \ea \right), ~~~~~[\rm{i = u,d}]~~
\label{e4}\ee 
where $D_i = |D_i|e^{i \sigma_i}$ and the elements of $M_i$ are assumed to
follow the hierarchical structure, e.g., $|D_i| \ll B_i \approx C_i < A_i$. 
The above matrices $M_i$  can be expressed as
\be M_i =P_i\bar M_i P_i^{\dagger}, \label{e5} \ee
where $\bar M_i$  the real matrices may be expressed as
\be \bar M_i  = \left(\ba{lll}
0 & |D_i| & 0 \\
| D_i| & C_i & B_i\\
0 & B_i & A_i\ea \right) \label{e6} 
\ee
and 
\be P_i  = diag \left(1, e^{-i\sigma_i}, e^{-i \sigma_i}
\right). \label{e7}\ee
The real matrices $\bar M_i$ can be diagonalized exactly by orthogonal
transformations, for example, 
\be \bar M_i  = O_i M_i^{diag}O_i^T \label{e8}\ee
where
\be 
M_i^{diag} = {\rm diag} \left({\rm m_1}, {\rm -m_2}, {\rm
m_3}\right) \label{e9} 
\ee
with subscripts 1, 2, 3 referring to u, c, t in `U' sector and d, s, b in
`D' sector. The details of diagonalizing matrix $O_i$ are given in
ref.\cite{r9}. \\ 
To facilitate comparison with FX calculations as well as for better
physical understanding of the structure of $V_{CKM}$, we present here the 
approximate form of $O_u$.  For example , by considering $m_u \ll m_c <
C_u < m_t$ as well as $m_d < m_s < C_d < m_b$, the structure for $O_u$ can 
be simplified and expressed as 
\be O_u \sim \left( \ba{ccc} 1 & -\sqrt{\frac{m_u}{m_c}} &
\frac{m_c}{m_t}\sqrt{\frac{m_u}{m_c} \left(\frac{m_c+C_u}{m_t-C_u}\right)}
\\
\sqrt{\frac{m_u}{m_c}\left(1-\frac{m_c+C_u}{m_t}\right)} &
\sqrt{1-\frac{C_u}{m_t}} & \sqrt{\frac{m_c+C_u}{m_t}}\\ - \sqrt{
\frac{m_u}{m_c} \left( \frac{m_c+C_u} {m_t} \right)} &
-\sqrt{\frac{m_c+C_u}{m_t}} & \sqrt{1- \frac {C_u}{m_t}}\ea \right) 
\label{e10}\ee
The matrix $O_d$  can be obtained  from $O_u$ simply  by  changing
$u\rightarrow d$, $c\rightarrow s$ and $t\rightarrow b$.\\ 
The mixing matrix $V_{CKM}$ in terms of $O_{u,d}$ can be expressed as
\be V_{CKM} = O_u^T P_{ud}O_d, \label{e11}\ee
where
\be P_{ud} = P_u^{\dagger}P_d  = diag \left( 1, e^{i \Delta \sigma},
e^{i \Delta \sigma}\right)~~~~ {\rm and}~~~ \Delta \sigma
=\sigma_u - \sigma_d. \label{e12} \ee
Using eq.(\ref{e10}) and retaining terms up to next-to-leading order,
eq.(\ref{e11}) can be simplified and written as 
\be V_{CKM} \cong \left(\ba{ccc}
1 & -\sqrt{\frac{m_d}{m_s}} & \frac{m_s}{m_b}
\sqrt { \frac {m_d} {m_s} \left( \frac {m_s+C_d} {m_b-C_d}\right) } +
\sqrt{\frac{m_u}{m_c}}g_2\\ 
-\sqrt{\frac{m_u}{m_c}} +\sqrt{\frac{m_d}{m_s}}g_1 & \sqrt{\frac{m_u
m_d}{m_c m_s}}+g_1 & g_2\\
\frac{m_c}{m_t}\sqrt{\frac{m_u}{m_c}\frac{m_c+C_u}{m_t-C_u}}
-\sqrt{\frac{m_d}{m_s}}g_2^{\prime} & -g_2 & g_1 \ea \right) 
\label{e13}\ee where
\be g_1 = \left(\sqrt{\left(1
-\frac{C_u}{m_t}\right)\left( 1- \frac{C_d}{m_b} \right)} +
\sqrt{\frac{m_c+C_u}{m_t}\frac{m_s+C_d}{m_t}}\right) e^{i \Delta \sigma}
\label{e14}\ee 
\be 
g_2 =\left(\sqrt{\frac{m_s+C_d}{m_b}\left(1-\frac{C_u}{m_t}\right)}-
\sqrt{\frac{m_c+C_u}{m_t}\left(1-\frac{C_d}{m_b}\right)}\right)
e^{i \Delta \sigma} \label{e15}\ee
and
\be  g^{\prime}_2 =
\left(\sqrt{\frac{m_s+C_d}{m_b}\left(1-\frac{C_u}{m_t}\right)}-
\sqrt{\frac{m_c+C_u}{m_t}\left(1-\frac{C_d+m_s}{m_b}\right)}\right)
e^{i \Delta \sigma} \label{e16}\ee
   The above expressions for $V_{CKM}$ are approximate, however, for
the purpose of calculations, we have employed exact expressions.\\
   After having calculated the $V_{CKM}$ elements, we  calculate  the
angles of the unitarity triangle related  to  the  decays
$B_d\rightarrow\pi\pi$,
$B_d\rightarrow D\pi$ and $B^0_d-\bar B^0_d$ mixing. The quantities
usually  discussed  are
$\sin2\alpha$, $\sin2\beta$ and $\sin2\gamma$ whose respective
relations to B decays  and
$V_{CKM}$  elements  are  detailed in ref.\cite{r10}. \\
   Before we present  our  results,  a  brief  discussion  about
various inputs which have gone into the analysis is in order. As a
first step, we have considered quark masses at 1  GeV  \cite{r11}, for
example, $m_u = .0051\pm.0015$~GeV, $m_d = .0089\pm.0026$~GeV,
$m_s = 0.175
\pm .055$~GeV, $m_c = 1.35\pm0.05$~GeV and $m_b= 5.3\pm0.1$~GeV.
 Unlike  FX
we have not studied the implications of the spread in mass values,
rather we have endeavored to understand the detailed  implications
of the variations of $C_u$ and $C_d$ on the CKM phenomenology.
Following
FX, to maximize CP violation, we have fixed $\Delta\sigma = 90^o$ .
Noting the
fact that the CKM matrix elements $|V_{us}|$ and $|V_{cb}|$ are well
known
and have weak dependence on $m_3$'s and $C_i$'s, we have restricted
the
parameter space by first reproducing $|V_{us}| = 0.22$ and $|V_{cb}|
\cong
0.038$ \cite{r12},\cite{r13}, ignoring the spread in the values of
$|V_{cb}|$ as the
calculated quantities hardly show any dependence on these. The
above value of $|V_{cb}|$, through the eqs. (\ref{e13}) and
(\ref{e15}), fixes the
values of $C_d$  for a given value of $C_u$.\\
   After having fixed the values of $|V_{us}|$ and $|V_{cb}|$, we have
calculated $|V_{ub}|$, $|V_{td}|$, $|V_{ts}|$  and  other
phenomenological
quantities related to $V_{CKM}$, for $m_t^{pole} = 175$~ GeV
(corresponding
to $m_t (1GeV) \approx 300$~ GeV ) and at different values of $R_t =
C_u /m_t$.
The  variations  w.r.t.  $m_t$ have  not  been  considered  as the
calculated  quantities  do  not  show  any  significant explicit
dependence on the present experimental range of
$m_t^{pole}$\cite{r6}.\\
   In tables $1$ and $2$ we have summarized the results of our
calculations for $\sin2\alpha$ being negative and  positive
respectively.
For the sake of uniformity,  we  have  presented  in  tables  the
ratios of $V_{CKM}$ matrix elements, e.g., $R_{ub} = |V_{ub}/V_{cb}|$
and $R_{td}=
|V_{td}/V_{cb}|$. A general survey of the tables brings out easily that
we are able to obtain $R_{ub}$ from 0.06-0.10 \cite{r10},\cite{r14} by
varying $R_t$
for $m_t^{pole} = 175$~GeV. This behaviour of $R_{ub}$ can be easily
checked
from the expressions of $V_{ub}$ and $V_{cb}$. Similarly the results
for
$|V_{td}|$ can encompass the presently expected range
\cite{r7}, \cite{r10}. Coming to the angles of unitarity triangle,
$\alpha$, $\beta$ and $\gamma$, we find that the present values are in
accordance  with
similar calculations by other authors \cite{r7},
\cite{r13},\cite{r15}.\\
   For the sake of brevity we have not included in the tables
our calculations regarding $V_{ts}$, $x_s$  and Jarlskog's CP violating
rephasing  invariant  parameter  J\cite{r16}.  However  a  few remarks
regarding these merit mention here. Interestingly we find that
$|V_{ts}| \leq |V_{cb}|$ for the entire range of $m_t^{pole}$ and $R_t$.
  This  is  in
accordance with expectations from the unitarity of $V_{CKM}$\cite{r10}.
A departure  from  the  above prediction would have important
implications for present texture specific mass matrices. Similarly
the parameter J and $B^0_s$-$\bar B^0_s$~mixing parameter $x_s$, though
not  shown
in the tables, have been  calculated  to  be  in  the  acceptable
ranges, e.g., $J = (1.3-2.4)\times10^{-5}$~\cite{r17}~ and~
$x_s$ = $8-15$ ($\sin2\alpha>0$),
$18-45$~ ($\sin2\alpha <0 $) for $x_d$ = $0.73$\cite{r10},
again  in  accordance  with
other similar calculations \cite{r7},\cite{r13},\cite{r15}.\\
   A careful study of the tables reveals several features  which
can be shown to be due to the presence of elements $C_u$ and $C_d$ in
the FX matrices.  One finds that there are distinct two ranges for
$V_{td}$, $\sin2\alpha$, $\sin2\beta$ ~and~ $\sin2\gamma$, although
in the case of  $\sin2\beta$  the
variations are not as much as in the case of
$\sin2\alpha$ and $\sin2\gamma$. It
is interesting to mention that $\sin2\beta$ does
not show much dependence
on $m_t$ or $C_u$ and $C_d$. The  narrow  range  of $\sin2\beta$,
despite
considerable variation of parameters, provides a  severe  testing
ground for the present texture $4$ zeros mass matrices.  The  effect
of the additional parameter is particularly  significant in  the
case of $|V_{td}/V_{cb}|$ which lies  in two ranges,  e.g., ~$.14-.22$~
and $.24-.31$, corresponding  respectively  to  lower  or  higher value
of $C_d$  generated by the eq.(\ref{e15}) for a given value of $C_u$.\\
   In order to have a better understanding of  the  significance of
our results, in fig. 1, we have shown the variation of
calculated values of $|V_{ub}|$ and $|V_{td}|$ as a function of $R_t$.
  It  is
interesting  to  mention  that  we  do  not  find  any  pronounced
dependence on $m_t$  for fixed $R_t$, however, the dependence on $C_u$
  and $C_d$  is considerable. The figure also shows two distinct ranges
  for
$V_{ub}$ and $V_{td}$ shown by  solid  and  dotted  lines
corresponding to
values of these in the tables  $1$  and  $2$  respectively.  The
reason for two branches is not difficult to  understand  when  one
realizes that fixing of $|V_{cb}|$  through eqs.(\ref{e13}) and
(\ref{e15}) leads to
two values for $C_d$  for a given value $C_u$. A detailed
investigation
of the exact expressions, e.g., eqs.(\ref{e13}), (\ref{e15})
and (\ref{e16}), brings
out clearly the contrasting behaviour of $V_{ub}$ and $V_{td}$.\\
   To understand the full significance of our  results  regarding
the two ranges for $V_{td}$, $\sin2\gamma$ etc., we  would  like  to
mention
that this can be linked to the ambiguity of the phase of  the  CKM
matrix, discussed in detail by Harris and Rosner \cite{r8}. To  explore
this further, we consider the exact standard  parameterization  of
the $V_{CKM}$ matrix\cite{r10}
\be V_{\nu} = \left(  \ba{ccc}
c_{12}c_{13} & s_{12}c_{13} & s_{13}e^{-i \delta} \\
-s_{12}c_{23}-c_{12}s_{23}s_{13}e^{i \delta} & c_{12} c_{23}-s_{12}
s_{23} s_{13}e^{i \delta} &
s_{23}c_{13} \\
s_{12}s_{23}-c_{12}c_{23}s_{13}e^{i \delta} &
-c_{12}s_{23}-s_{12}c_{23}s_{13}e^{i \delta} &
c_{23}c_{13} \ea\right)
\label{e17} \ee
where $c_{ij}  = \cos\theta_{ij}$, $s_{ij} = \sin\theta_{ij}$~~{\rm
   and}~~ $\delta$ corresponds to $\delta_{13}$ of the
standard parameterization.\\
  In this parameterization $\delta$ is simply equal  the  angle
  $\gamma$  of
unitarity triangle. As the value of $\beta$ does not  show  much
variation, therefore, the behaviour of $\delta$ is very much related to
the variations of angles $\alpha$ and $\gamma$. As  has  been
mentioned in  the
tables, the present data allows an ambiguity in the sign of
$\sin2\gamma$
which easily translates to the ambiguity of the quadrant of $\delta$
\cite{r8}.
For having a feeling of this ambiguity in the $V_{CKM}$ matrix
elements,
 one has to closely analyze  the  exact  expression  for $V_{td}$ in
eqs.(\ref{e13}) and (\ref{e16}) along with $V_{cb}$ in eqs.
(\ref{e13}) and (\ref{e15}). \\
   From the tables and the expression (\ref{e17}) one can  easily  find
the range of $\delta$ predicted by the present set of mass matrices, as
follows
\be 22^o \leq \delta \leq 45^o~~~{\rm and}~~~ 95^o \leq \delta\leq
130^o.\label{e18} \ee
in good overlap with other similar calculations\cite{r7}.
   To conclude, FX mass matrices have  been  investigated  in
the context of CKM phenomenology. In particular, we have examined
in detail the implications of additional elements introduced by FX
in comparison with the earlier Fritzsch mass matrices. Apart  from
obtaining the non-leading order corrections to eqs.
(\ref{e1})-(\ref{e3}),
 we have seen that the maximality of CP-violation as enunciated by  FX
is in tune with the present data. It is very striking to note that
the variation of $V_{td}$, due to the additional elements $C_{u,d}$,
 leads
to a range which is  very  much  in  agreement  with  the  results
expected from the recently improved calculations
of $B^0-\bar B^0$ mixing
\cite{r7} for a fairly broad range of  parameters  considered
here.
The role played by $C_u$  and $C_d$  in fitting the recent
data pertaining
to CKM matrix elements needs to be highlighted as when  either  of
them is zero the full range of data pertaining to $V_{ub}$ and
$V_{td}$ can
not be fitted \cite{r18}. A precise measurement of angle $\beta$,
 through the
decay $B\rightarrow \psi K^0_s$, and $V_{td}$ would certainly help in
establishing  the
validity of present set of  mass  matrices.  Further,  in  the
language of Ramond et al. \cite{r4} it seems that present
data  favours
texture $4$ zeros mass matrices.\\
\vskip .2truecm
The authors gratefully acknowledge a few useful discussions with
Prof. M. P. Khanna. PSG would like to thank the chairman,
Department of
Physics for providing facilities to work in the department
as well as
principal of his college for his kind cooperation.\\

\newpage
\begin{table}
\label{1a}
\begin{tabular}{|c|c|c|c|c|c|}
\hline
$R_t$ & $R_{ub}$ & $R_{td}$ & $S_{2 \alpha}$ & $S_{2 \beta}$ & $ S_{2
\gamma}$\\
\hline
0.00 & 0.06 & 0.21 & -0.18 & 0.52 & 0.66\\
0.02 & 0.07 & 0.20 & -0.58 & 0.52 & 0.95\\
0.04 & 0.07 & 0.18 & -0.88 & 0.53 & 1.00\\
0.06 & 0.08 & 0.18 & -0.91 & 0.54 & 0.99\\
0.08 & 0.08 & 0.17 & -0.96 & 0.54 & 0.96\\
0.10 & 0.08 & 0.16 & -0.99 & 0.54 & 0.91 \\
0.12 & 0.09 & 0.15 & -1.00 & 0.55 & 0.87\\
0.14 & 0.09 & 0.15 & -1.00 & 0.55 & 0.82\\
0.16 & 0.10 & 0.14 & -1.00 & 0.56 & 0.75\\
\hline
\end{tabular}
\caption{Calculated values of $R_{ub}$, $R_{td}$,
$S_{2 \theta}( =\sin2\theta,~~\theta=\alpha, \beta,\gamma$)
for $m_t^{pole}= 175$~ GeV and for different $R_t$ values
when $|V_{cb}| = 0.038$ with $S_{2 \alpha}$ taking negative values.}
\end{table}
\begin{table}
\label{1b}
\begin{tabular}{|c|c|c|c|c|c|}
\hline
$R_t$ & $R_{ub}$ & $R_{td}$ & $S_{2 \alpha}$ & $S_{2 \beta}$ & $ S_{2
\gamma}$\\
\hline
0.00 & 0.06 & 0.24 & 0.61 & 0.51 & -0.09\\
0.02 & 0.07 & 0.26 & 0.90 & 0.50 & -0.47\\
0.04 & 0.08 & 0.27 & 0.98 & 0.50 & -0.70\\
0.06 & 0.08 & 0.28 & 1.00 & 0.50 & -0.80\\
0.08 & 0.09 & 0.29 & 1.00 & 0.49 & -0.88\\
0.10 & 0.09 & 0.30 & 0.98 & 0.49 & -0.93 \\
0.12 & 0.10 & 0.30 & 0.96 & 0.49 & -0.96\\
\hline
\end{tabular}
\caption{Same as in table 2 with $S_{2 \alpha}$ taking positive
values.}
\end{table}
\setlength{\unitlength}{0.240900pt}
\ifx\plotpoint\undefined\newsavebox{\plotpoint}\fi
\sbox{\plotpoint}{\rule[-0.200pt]{0.400pt}{0.400pt}}%
\begin{picture}(1049,629)(0,0)
\font\gnuplot=cmtt10 at 12pt
\gnuplot
\sbox{\plotpoint}{\rule[-0.200pt]{0.400pt}{0.400pt}}%
\put(120.0,31.0){\rule[-0.200pt]{0.400pt}{142.372pt}}
\put(120.0,53.0){\rule[-0.200pt]{4.818pt}{0.400pt}}
\put(108,53){\makebox(0,0)[r]{0.06}}
\put(985.0,53.0){\rule[-0.200pt]{4.818pt}{0.400pt}}
\put(120.0,140.0){\rule[-0.200pt]{4.818pt}{0.400pt}}
\put(108,140){\makebox(0,0)[r]{0.1}}
\put(985.0,140.0){\rule[-0.200pt]{4.818pt}{0.400pt}}
\put(120.0,228.0){\rule[-0.200pt]{4.818pt}{0.400pt}}
\put(108,228){\makebox(0,0)[r]{0.14}}
\put(985.0,228.0){\rule[-0.200pt]{4.818pt}{0.400pt}}
\put(120.0,316.0){\rule[-0.200pt]{4.818pt}{0.400pt}}
\put(108,316){\makebox(0,0)[r]{0.18}}
\put(985.0,316.0){\rule[-0.200pt]{4.818pt}{0.400pt}}
\put(120.0,403.0){\rule[-0.200pt]{4.818pt}{0.400pt}}
\put(108,403){\makebox(0,0)[r]{0.22}}
\put(985.0,403.0){\rule[-0.200pt]{4.818pt}{0.400pt}}
\put(120.0,491.0){\rule[-0.200pt]{4.818pt}{0.400pt}}
\put(108,491){\makebox(0,0)[r]{0.26}}
\put(985.0,491.0){\rule[-0.200pt]{4.818pt}{0.400pt}}
\put(120.0,578.0){\rule[-0.200pt]{4.818pt}{0.400pt}}
\put(108,578){\makebox(0,0)[r]{0.3}}
\put(985.0,578.0){\rule[-0.200pt]{4.818pt}{0.400pt}}
\put(120.0,31.0){\rule[-0.200pt]{0.400pt}{4.818pt}}
\put(120,10){\makebox(0,0){0}}
\put(120.0,602.0){\rule[-0.200pt]{0.400pt}{4.818pt}}
\put(268.0,31.0){\rule[-0.200pt]{0.400pt}{4.818pt}}
\put(268,10){\makebox(0,0){0.03}}
\put(268.0,602.0){\rule[-0.200pt]{0.400pt}{4.818pt}}
\put(415.0,31.0){\rule[-0.200pt]{0.400pt}{4.818pt}}
\put(415,10){\makebox(0,0){0.06}}
\put(415.0,602.0){\rule[-0.200pt]{0.400pt}{4.818pt}}
\put(563.0,31.0){\rule[-0.200pt]{0.400pt}{4.818pt}}
\put(563,10){\makebox(0,0){0.09}}
\put(563.0,602.0){\rule[-0.200pt]{0.400pt}{4.818pt}}
\put(710.0,31.0){\rule[-0.200pt]{0.400pt}{4.818pt}}
\put(710,10){\makebox(0,0){0.12}}
\put(710.0,602.0){\rule[-0.200pt]{0.400pt}{4.818pt}}
\put(858.0,31.0){\rule[-0.200pt]{0.400pt}{4.818pt}}
\put(858,10){\makebox(0,0){0.15}}
\put(858.0,602.0){\rule[-0.200pt]{0.400pt}{4.818pt}}
\put(1005.0,31.0){\rule[-0.200pt]{0.400pt}{4.818pt}}
\put(1005,10){\makebox(0,0){0.18}}
\put(1005.0,602.0){\rule[-0.200pt]{0.400pt}{4.818pt}}
\put(120.0,31.0){\rule[-0.200pt]{213.196pt}{0.400pt}}
\put(1005.0,31.0){\rule[-0.200pt]{0.400pt}{142.372pt}}
\put(120.0,622.0){\rule[-0.200pt]{213.196pt}{0.400pt}}
\put(-40,326){\makebox(0,0){$R_{ij}$}}
\put(562,-22){\makebox(0,0){$R_t$}}
\put(120.0,31.0){\rule[-0.200pt]{0.400pt}{142.372pt}}
\put(900,550){\makebox(0,0)[r]{$R_{ub1}$}}
\put(120,53){\usebox{\plotpoint}}
\multiput(120.00,53.58)(4.821,0.499){171}{\rule{3.944pt}{0.120pt}}
\put(925,550){\raisebox{-.8pt}{\makebox(0,0){$\Diamond$}}}
\put(120,53){\raisebox{-.8pt}{\makebox(0,0){$\Diamond$}}}
\put(956,140){\raisebox{-.8pt}{\makebox(0,0){$\Diamond$}}}
\put(900,500){\makebox(0,0)[r]{$R_{ub2}$}}
\put(120,57){\usebox{\plotpoint}}
\multiput(120,57)(20.583,2.673){32}{\usebox{\plotpoint}}
\put(759,140){\usebox{\plotpoint}}
\put(925,500){\makebox(0,0){$+$}}
\put(120,57){\makebox(0,0){$+$}}
\put(759,140){\makebox(0,0){$+$}}
\sbox{\plotpoint}{\rule[-0.400pt]{0.800pt}{0.800pt}}%
\put(900,450){\makebox(0,0)[r]{$R_{td1}$}}
\put(120,381){\usebox{\plotpoint}}
\multiput(120.00,379.09)(2.742,-0.501){299}{\rule{4.571pt}{0.121pt}}
\put(925,450){\raisebox{-.8pt}{\makebox(0,0){$\Box$}}}
\put(120,381){\raisebox{-.8pt}{\makebox(0,0){$\Box$}}}
\put(956,228){\raisebox{-.8pt}{\makebox(0,0){$\Box$}}}
\sbox{\plotpoint}{\rule[-0.500pt]{1.000pt}{1.000pt}}%
\put(900,400){\makebox(0,0)[r]{$R_{td2}$}}
\put(120,447){\usebox{\plotpoint}}
\multiput(120,447)(20.185,4.833){32}{\usebox{\plotpoint}}
\put(759,600){\usebox{\plotpoint}}
\put(925,400){\makebox(0,0){$\times$}}
\put(120,447){\makebox(0,0){$\times$}}
\put(759,600){\makebox(0,0){$\times$}}
\end{picture}
\vskip .2truecm
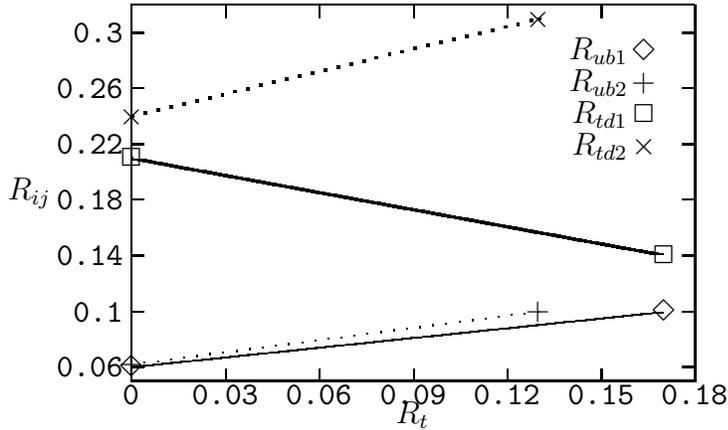
{Fig.1 \hskip .2truecm Variation of calculated values of
$|V_{ub}/V_{cb}|$ and
$|V_{td}/V_{cb}|$ w.r.t. $R_t$ ($R_{ij} =|V_{ij}/V_{cb}|$, ij = ub, td
),
suffixes 1 and 2
corresponding to tables 1 and 2 respctively.}
\end{document}